\documentclass[pre,a4paper,amsfonts,amssymb,onecolumn,showpacs,showkeys,nofootinbib,floats]{revtex4}

\usepackage{amsfonts}
\usepackage{amsmath}
\usepackage{amssymb}
\usepackage{graphicx}

\begin{document}

\title{On statistical properties of traded volume in financial markets}

\author{J.~de~Souza}
\email[e-mail address: ]{jeferson@cbpf.br}
\affiliation{Centro Brasileiro de Pesquisas F\'{\i}sicas, 150,
22290-180, Rio de Janeiro - RJ, Brazil}
\author{L.~G.~Moyano}
\email[e-mail address: ]{moyano@cbpf.br}
\affiliation{Centro Brasileiro de Pesquisas F\'{\i}sicas, 150,
22290-180, Rio de Janeiro - RJ, Brazil}
\author{S.~M.~Duarte~Queir\'{o}s}
\email[e-mail address (Corresponding author): ]{sdqueiro@cbpf.br}
\affiliation{Centro Brasileiro de Pesquisas F\'{\i}sicas, 150,
22290-180, Rio de Janeiro - RJ, Brazil}

\date{\today}

\begin{abstract}
In this article we study the dependence degree of the traded volume of
the Dow Jones 30 constituent equities
by using a nonextensive generalised form of the Kullback-Leibler
information measure. Our results
show a slow decay of the dependence degree as a function of the lag.
This feature is compatible with the existence
of non-linearities in this type time series. In addition, we introduce a
dynamical mechanism whose associated
stationary probability density function (PDF) presents a good agreement
with the empirical results.
\end{abstract}



\maketitle

\section{Introduction}
The study of complexity, in particular within financial systems, has
become one of the main focus of interest
in statistical physics~\cite{gm-ct}.
In fact, several statistical properties verified in financial
observables, {\it e.g.}, relative
price changes (\textit{the return}) and returns standard deviation
(\textit{the volatility}), have enabled the establishment
of new models which characterise systems ever
better~\cite{stanley-bouchaud-voit}.
Along with the previous two quantities, another key observable is the
number of stocks of a certain company traded in
a given period of time, \textit{the traded volume},
$v$. In this article we analyse the dependence degree of $1$-minute
traded volume time series, $V(t)$, of the constituents
of the Dow Jones Industrial Average $30$ index (DJ30), between the
$1^{st}$ of July 2004
and the $31^{st}$ of December 2004. We introduce also a dynamical
mechanism that
provides the same stationary PDF~\cite{gopi-volumes,obt,epl-volumes}. In
order to avoid spurious features, we have removed
intra-day pattern of the original time series and normalised each
element of the series
by its mean value defining the normalised traded volume, 
$
v\left( t\right) =\frac{V^{\prime }\left( t\right) }{\left\langle V^{\prime
}\left( t\right) \right\rangle ,}
$
where $V^{\prime }\left( t\right) =\frac{V\left( t\right) }{\Xi \left(
t^{\prime } \right) }
$, $\Xi \left( t^{\prime }\right) =\frac{\sum\limits_{i=1}^{N}V\left(
t_{i}^{\prime }\right) }{N}$
and $\langle \ldots \rangle $ is defined as the average over time
($t^{\prime }$ represents
the intra-day time and $i$ the day).

\section{Dependence degree}
Discrimination between two hypothesis, {\it consistent testing}, is
ubiquitous in science.
Examples are the stationary/non-stationary character of
time series or the dependence degree between its elements. Concerning
the latter, the most widely applied measure
of ``dependence'' between variables is the correlation function
mathematically defined as,
$$C\left[ v\left( t\right) ,v\left( t+\tau \right) \right]
=\frac{\left\langle
v\left( t\right) \,v\left( t+\tau \right) \right\rangle -\left\langle
v\left( t\right) \right\rangle ^{2}}{\left\langle v\left( t\right)
^{2}\right\rangle -\left\langle v\left( t\right) \right\rangle ^{2}}.$$
Since the correlation function is basically a normalised covariance (or
the second cumulant of the stochastic
process), it will only be a suitable statistical procedure for linear
correlations or correlations that can be
written in a linear way. In other words, the correlation function is not
able to determine conveniently non-linearities
in a given group of data. Aiming to consistently test the dependence or
independence of stochastic variables it was recently defined
a dependence measure that has been able to evaluate non-linearities, for
instance, in daily return time
series~\cite{smdq-quantf} and GARCH processes~\cite{garch} for which the
correlation function gives zero value.

So, let us start by defining our dependence measure as the {\it
non-extensive generalised mutual information measure},
\begin{equation*}
I_{q^{\prime }}=-\int p\left( y\right) \,\ln _{q^{\prime }}\frac{p^{\prime
}\left( y\right) }{p\left( y\right) }\,dy
\end{equation*}
where $\ln _{q^{\prime }}\left( y\right) =\frac{y^{q^{\prime
}-1}-1}{q^{\prime }-1}$
($\ln _{q^{\prime }}\left( y\right) = \ln _{1}\left( y\right)$),
which emerged within the non-extensive formalism based on Tsallis
entropy~\cite{ct}.
For $q^{\prime }=1$, it is equivalent to the Kullback-Leibler information
gain~\cite{ct-kl,kl}.

Let us now assume that $y$ is a two-dimensional random
variable $y=\left( x,z\right) $. We can quantify the degree dependence
between $x$ and $z$ by computing $I_{q^{\prime }}$ for $p\left(
x,z\right) $ and
$p^{\prime }\left( x,z\right) =p_{_{1}}\left( x\right) \,p_{_{2}}\left(
z\right) $, where $p_{\ldots }(\ldots )$ represents the marginal
probability.
For this case, $I_{q^{\prime }}$ presents both a lower bound and an
upper bound.
The former, $I_{q^{\prime }}^{MIN}=0$, corresponds to {\it total
independence} between
$x$ and $z$, i.e. $  p \left( x,z\right)= p^{\prime }\left( x,z\right)$.
The latter, $I_{q^{\prime }}^{MAX}$, represents a {\it one-to-one
dependence} between variables and is given by,
\begin{equation*}
\begin{array}{ccc}
I_{q^{\prime }}^{MAX} & = & -\int \int p\left( x,z\right) \,\left[ \ln
_{q^{\prime }}p_{1}\left( x\right) +\right.  \\
&  & \left. (1-q)\,\ln _{q^{\prime }}\,p_{_{1}}\left( x\right) \,\,\ln
_{q^{\prime }}\,p_{_{2}}\left( z\right) \right] \,dx\,dz.%
\end{array}
\end{equation*}
 From these two extreme values, it is then possible to define a
normalised measure,
$$R_{q^{\prime }}=\frac{I_{q^{\prime }}^{{}}}{I_{q^{\prime }}^{MAX}}\in
\left[ 0,1\right],$$
 which has an {\it optimal} index, $q^{op}$ (where the prime was
suppressed for clarity).

This index is optimal in the sense that the gradient of the measure $R$
is most sensitive and hence
most capable of determine variations in the dependence among the
variables. Moreover, it is optimal
because its two extreme values are associated to full dependence and
full independence between $x$ and $z$.
Analytically, it is determined by the inflection point of  $R_{q^{\prime
}}$
\textit{vs} $q^{\prime }$ curves. For one-to-one dependence we have
$q^{op}=0$, and
$q^{op}= \infty$ for total independence (see reference \cite{ct-kl} for
a detailed discussion).

\begin{figure}[tbp]
\begin{center}
\includegraphics[width=0.45\columnwidth,angle=0]{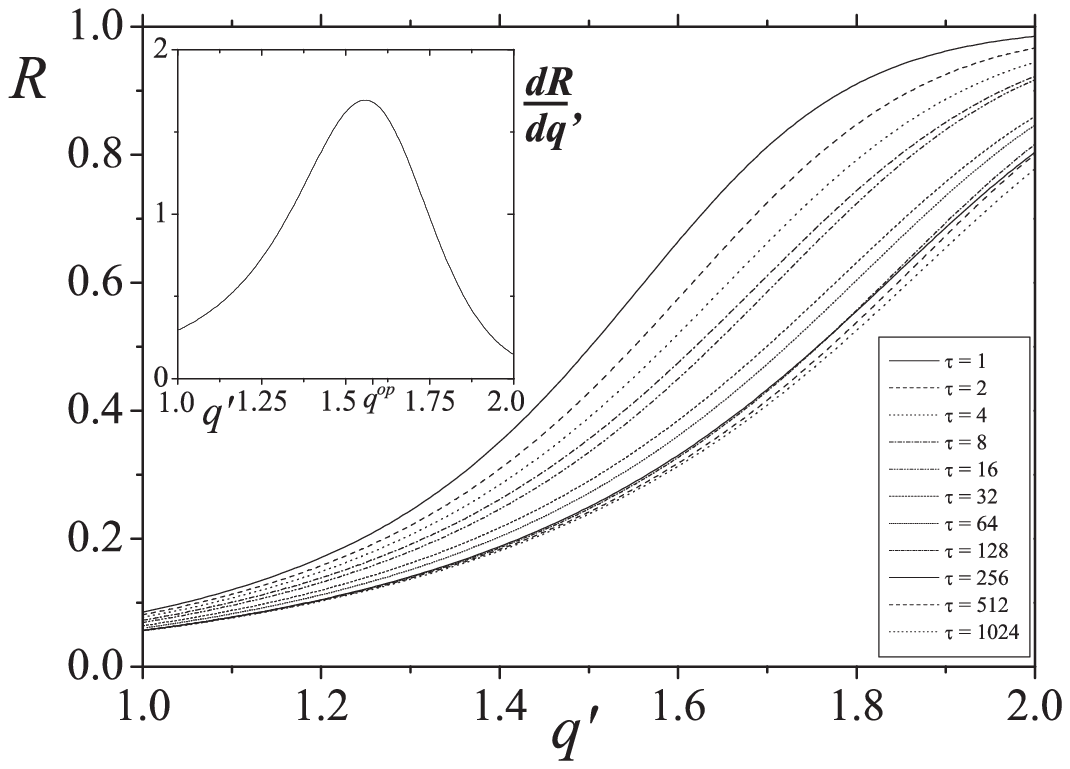}
\includegraphics[width=0.45\columnwidth,angle=0]{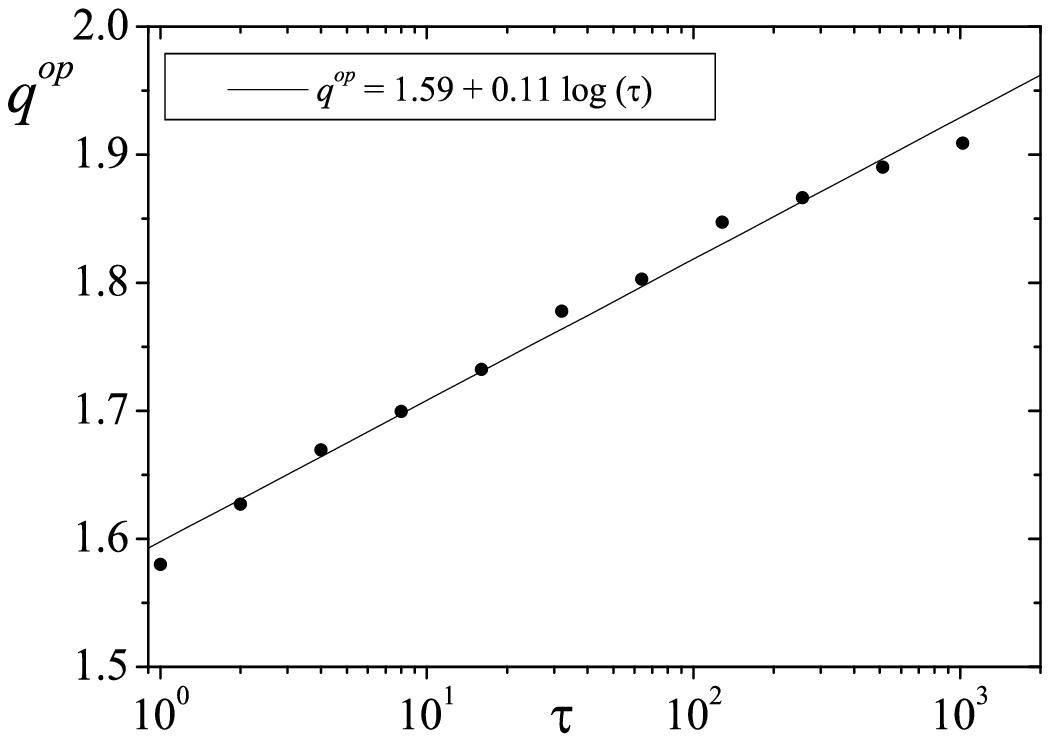}
\end{center}
\caption{ Left: Normalised generalised Kullback-Leibler measure,
$R_{q^{\prime }}$,
{\it vs.} entropic index, $q^{\prime }$, for the International Business
Machines (IBM).
The inset shows, as mere illustration, the derivative of $R$ in respect
to $q^{\prime }$ for $\tau = 1$ . The maximum corresponds to
$q^{op}=1.58$. Right: The symbols represent the dependence
degree, $q^{op}$, {\it vs.} $\tau $ (in minutes) averaged over the $30$
time series.
The line represents a fitting logarithmic function ($q^{op} = 1.59 +
0.11 \log (\tau )$) (the correlation coefficient is
$0.9944$) pointing up the slow increase of $q^{op}$.} \label{qopt}
\end{figure}

\begin{figure}[tbp]
\begin{center}
\includegraphics[width=0.45\columnwidth,angle=0]{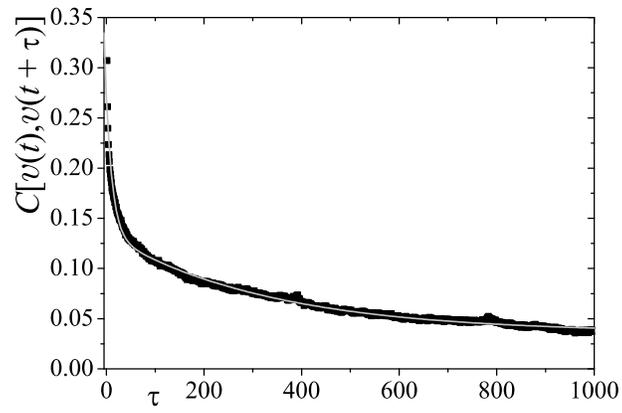}
\end{center}
\caption{Left: Symbols represent the average correlation function for
the 30 time series
analysed and the line represents a double exponential fit with
characteristic times of  
$\gamma ^{-1}=13$ and $T = 332$ yielding a ratio about $25$ between the
two time scales  
Eq.~(\ref{correlation}) ($R^2 = 0.991$ and $\chi ^2 = 9 \times 10^{-6}$
and time in minutes).
}
\label{cor}
\end{figure}

We have computed $R_{q^{\prime }}$ for all time series with $x = v(t)$,
$z = v(t+ \tau )$, where $\tau $
represents the lag. A typical example is presented in Fig.~\ref{qopt}
(left panel).
Analysing the behaviour of $q^{op}$ as a function of $\tau $, we have
observed a slow
increase of $q^{op}$, i.e., a slow decrease in the dependence degree
between variables as it is visible in
Fig.~\ref{qopt} (left panel). Our result reveals the existence of
significant non-linear dependences which seem to
be present even for large times. In Fig.~\ref{cor} it is possible to see
that the correlation value between $\tau = 1$ and $\tau = 1000$
diminishes around $80 \% $ while the $q^{op}$ value between $\tau =
1024$ and
$\tau = 1$ only reduces in $20 \%$ (approximately), i.e., a decrease in
the dependence
degree in the same amount.

\begin{table}
\caption{Obtained values from: PDF fitting ($q$, $\theta $ and  $\alpha
$) and from correlation analysis ($\gamma \, T$).}
\begin{tabular}{lcccc}
& ${\scriptsize q}$ & ${\scriptsize \theta }$ & ${\scriptsize \alpha }$ & $%
{\scriptsize T\,\gamma }$ \\ \hline\hline
{\scriptsize AA} & {\scriptsize 1.19} & {\scriptsize 8.81\thinspace } &
{\scriptsize 2.67\thinspace } & {\scriptsize 29} \\
{\scriptsize AIG} & {\scriptsize 1.22} & {\scriptsize 4.32} & {\scriptsize %
1.84} & {\scriptsize 34} \\
{\scriptsize AXP} & {\scriptsize 1.21} & {\scriptsize 6.51} & {\scriptsize %
2.06} & {\scriptsize 26} \\
{\scriptsize BA} & {\scriptsize 1.18} & {\scriptsize 10.67} & {\scriptsize %
2.95} & {\scriptsize 24} \\
{\scriptsize C} & {\scriptsize 1.15} & {\scriptsize 9.20} & {\scriptsize
3.18%
} & {\scriptsize 25} \\
{\scriptsize CAT} & {\scriptsize 1.20} & {\scriptsize 7.49} & {\scriptsize %
2.32} & {\scriptsize 13} \\
{\scriptsize DD} & {\scriptsize 1.20} & {\scriptsize 7.33} & {\scriptsize %
2.26} & {\scriptsize 53} \\
{\scriptsize DIS} & {\scriptsize 1.21} & {\scriptsize 7.29} & {\scriptsize %
2.19} & {\scriptsize 20} \\
{\scriptsize GE} & {\scriptsize 1.17} & {\scriptsize 8.31} & {\scriptsize %
2.75} & {\scriptsize 33} \\
{\scriptsize GM} & {\scriptsize 1.21} & {\scriptsize 8.14} & {\scriptsize %
2.46} & {\scriptsize 29} \\
{\scriptsize HD} & {\scriptsize 1.17} & {\scriptsize 8.76} & {\scriptsize %
2.84} & {\scriptsize 27} \\
{\scriptsize HON} & {\scriptsize 1.19} & {\scriptsize 9.06} & {\scriptsize %
2.67} & {\scriptsize 70} \\
{\scriptsize HPQ} & {\scriptsize 1.19} & {\scriptsize 8.55} & {\scriptsize %
2.64} & {\scriptsize 28} \\
{\scriptsize IBM} & {\scriptsize 1.14} & {\scriptsize 12.36} &
{\scriptsize %
3.70} & {\scriptsize 41} \\
{\scriptsize INTC} & {\scriptsize 1.20} & {\scriptsize 4.22} &
{\scriptsize %
1.70} & {\scriptsize 25} \\
{\scriptsize JNJ} & {\scriptsize 1.17} & {\scriptsize 8.55} & {\scriptsize %
2.91} & {\scriptsize 11} \\
{\scriptsize JPM} & {\scriptsize 1.17} & {\scriptsize 9.14} & {\scriptsize %
2.92} & {\scriptsize 22} \\
{\scriptsize KO} & {\scriptsize 1.19} & {\scriptsize 7.88} & {\scriptsize %
2.61} & {\scriptsize 26} \\
{\scriptsize MCD} & {\scriptsize 1.21} & {\scriptsize 7.48} & {\scriptsize %
2.30} & {\scriptsize 30} \\
{\scriptsize MMM} & {\scriptsize 1.19} & {\scriptsize 7.14} & {\scriptsize %
2.33} & {\scriptsize 23} \\
{\scriptsize MO} & {\scriptsize 1.18} & {\scriptsize 7.73} & {\scriptsize %
2.66} & {\scriptsize 12} \\
{\scriptsize MRK} & {\scriptsize 1.25} & {\scriptsize 1.24} & {\scriptsize %
0.61} & {\scriptsize 21} \\
{\scriptsize MSFT} & {\scriptsize 1.22} & {\scriptsize 4.57} &
{\scriptsize %
1.62} & {\scriptsize 23} \\
{\scriptsize PFE} & {\scriptsize 1.18} & {\scriptsize 6.31} & {\scriptsize %
2.44} & {\scriptsize 33} \\
{\scriptsize PG} & {\scriptsize 1.16} & {\scriptsize 8.94} & {\scriptsize %
2.99} & {\scriptsize 23} \\
{\scriptsize SBC} & {\scriptsize 1.19} & {\scriptsize 8.62} & {\scriptsize %
2.57} & {\scriptsize 25} \\
{\scriptsize UTX} & {\scriptsize 1.14} & {\scriptsize 18.47} &
{\scriptsize %
4.71} & {\scriptsize 32} \\
{\scriptsize VZ} & {\scriptsize 1.17} & {\scriptsize 8.83} & {\scriptsize %
2.84} & {\scriptsize 34} \\
{\scriptsize WMT} & {\scriptsize 1.16} & {\scriptsize 10.24} &
{\scriptsize %
3.23} & {\scriptsize 30} \\
{\scriptsize XOM} & \multicolumn{1}{l}{\scriptsize 1.15} &
\multicolumn{1}{l}%
{\scriptsize 11.45} & \multicolumn{1}{l}{\scriptsize 3.50} &
\multicolumn{1}{l}{\scriptsize 31} \\ \hline
\end{tabular}
\end{table}

\section{A possible dynamical model for traded volumes}
\label{mod-sec}
The non-linear character of a time series manifests on the exhibition of
(asymptotic)
power-law behaviour of the stochastic variable (stationary) PDF. This
power-law-like behaviour of the PDF was
also verified for traded volume time series~\cite{gopi-volumes,obt}. In
order to describe a possible dynamical mechanism for this
observable, let us suppose that the traded volume of an equity is
described by the following stochastic differential equation,
\begin{equation}
dv=-\gamma (v-\frac{\omega }{\alpha })\,dt+\sqrt{2\frac{\,\gamma
\,}{\alpha }%
}v\,dW_{t},  \label{feller}
\end{equation}
where $W_{t}$ is a regular Wiener process following a normal distribution
and $v\geq 0$. The right-hand side of Eq.~(\ref{feller}) may be
interpreted as follows:
the deterministic term represents a natural mechanism of the system
which aims to keep the traded volume at some ``normal'' value, $\omega
/\alpha $ with a relaxation time
of order of $\gamma ^{-1}$. The stochastic term mimics the microscopic
effects on the evolution of $v$,
just like a multiplicative noise used to replicate intermittent
processes. This dynamics and the
corresponding Fokker-Planck equation~\cite{risken} leads to an inverted
Gamma stationary distribution,
\begin{equation}
f\left( v\right) =\frac{\,\,1}{\omega \,\Gamma \left[ \alpha +1\right] }%
\left( \frac{v}{\omega }\right) ^{-\alpha -2} \, \exp \left[
-\frac{\,\omega }{v}%
\right] .  \label{f-v}
\end{equation}
Consider now, in the same lines of Beck and Cohen
superstatistics~\cite{beck-cohen},
that instead of constant, $\omega $ is a time dependent
quantity which evolves on a time scale $T$ larger than the time scale
$\gamma ^{-1}$ required by Eq.~(\ref{feller}) to reach stationarity.
This time dependence is, in the present model,
associated to changes in the volume of activity (number of traders that
performed transactions)~\cite{epl-volumes}.
Furthermore, if we assume that $\omega $ follows a Gamma PDF,
\begin{equation}
P\left( \omega \right) =\frac{1}{\lambda \Gamma \left[ \delta \right] }%
\left( \frac{\omega }{\lambda }\right) ^{\delta -1}\exp \left[
-\frac{\omega
}{\lambda \,}\right] ,  \label{p-omega}
\end{equation}
the long-term distribution of $v$ will be given by $p\left( v\right) =\int
f\left( v\right) \,P\left( \omega \right) \,d\omega $ which yields,
\begin{equation}
p\left( v\right) =\frac{1}{Z}\left( \frac{v}{\theta }\right) ^{-\alpha
-2}\exp _{q}\left[ -\frac{\theta }{v}\right]  \label{p-v}
\end{equation}
where $\lambda =\theta \left( q-1\right) $, $\delta
=\frac{1}{q-1}-\alpha -1$
and $\exp _{q}\left[ x\right] \equiv \left[ 1+\left( 1-q\right) \,x\right]
^{1/\left( 1-q\right) }$ is the $q$-exponential function, the inverse
function of $\ln _{q}\left( y\right) $ ($\exp _{1}\left[ x\right]
=e^{x}$)~\cite{ct},
$Z$ being the normalisation constant.

\begin{figure}[tbp]
\begin{center}
\includegraphics[width=0.45\columnwidth,angle=0]{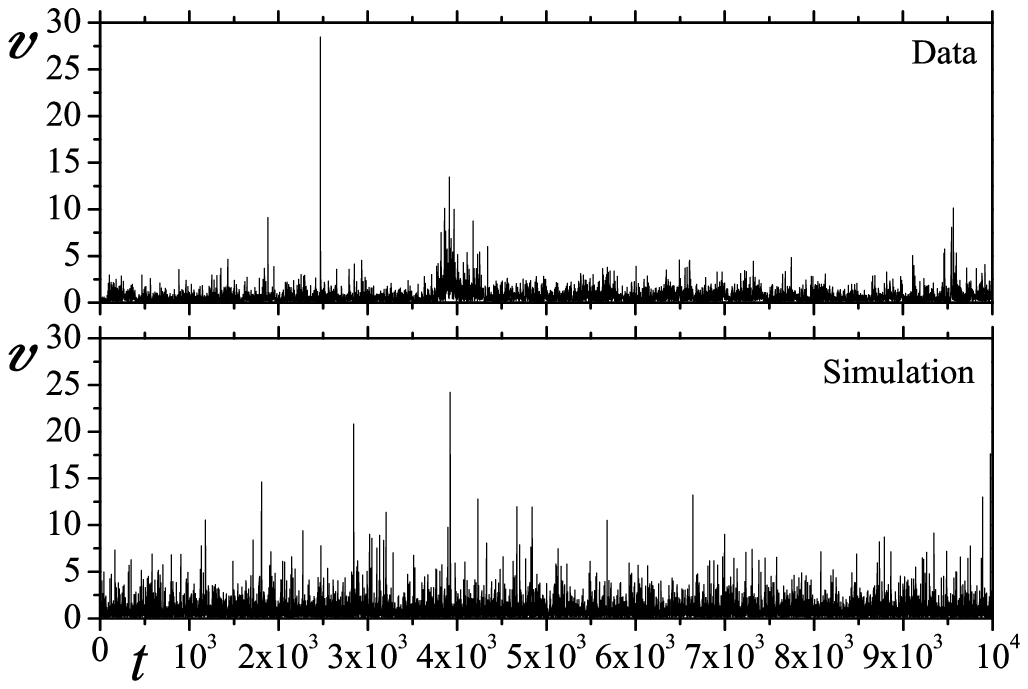}
\includegraphics[width=0.45\columnwidth,angle=0]{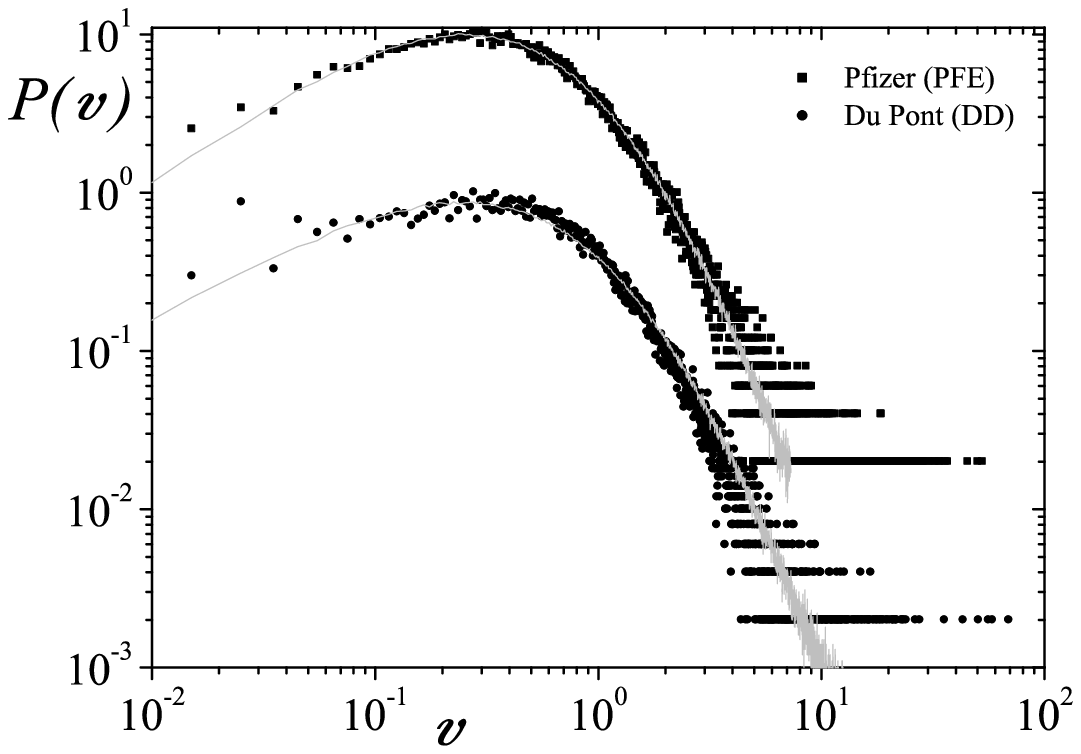}
\end{center}
\caption{Left: (Upper panel) Excerpt from the analysed Pfizer time
series; (Lower panel) Excerpt from the time
series generated to mimic Pfizer using the values presented in Tab.~I.
($t$ in minutes)
Right: Symbols represent the empirical PDF for Pfizer (shifted by a
factor of $10$) and
Du Pont normalised traded volume time series, which correspond to the
best ($R^2 = 0.9953$ and $\chi ^2 = 0.0002$)
($R^2 = 0.9763$ and $\chi ^2 = 0.001$) and worst fits, respectively.
The lines correspond to simulation using the values presented in Tab.~I.}
\label{pdf}
\end{figure}

This approach is probabilistically equivalent to the one
in~\cite{epl-volumes,nota-ct},
but it is more realistic concerning the dependence on $v$ of the
Kramers-Moyal moments.
In other words, this model is, \textit{in principle}, a better dynamical
approach.
In regard of the measured values of $q$, $\theta $, $\alpha $ in Tab.~I,
we verify that they are enclosed within
a small interval in the $q$ values, $1.19\pm 0.02$ (close to
$\frac{6}{5}$) and presents wider intervals
for the other parameters, $\alpha =2.63 \pm 0.48$ and $\theta =8.31\pm
1.86$.
In Fig.~\ref{pdf} we present the best (Pfizer, PFE) and the worst (Du
Pont, DD) fits.

With the $\alpha $, $\theta $ and $q$ fitting values in Tab.~I we have
generated a set of time series aiming to test the validity
of our approach. For the evaluation of the time scales $\gamma ^{-1}$
and $T$, we have considered the simplest approach, i.e.,
the ratio between the two time scales which describe the CF for traded
volume. See equity values of $\gamma \, T$
in Tab.~I. As can be seen from Fig.~\ref{cor}, there is a fast decay of
the CF,
related to local equilibrium, and then a much slower decay for larger
times that are due to a
slow decay of correlations in $\omega $, i.e.,
\begin{equation}
C\left[ v\left( t\right) ,v\left( t+\tau \right) \right]
=C_{1} \, e^{-\gamma \,\tau }+C_{2} \, e^{-\,\tau /T}.  \label{correlation}
\end{equation}
 This slow decay is consistent with a slow dynamics of $\omega $,
necessary condition for the appliance
of a superstatistical model. In our numerical calculations we have
defined time in $\gamma ^{-1}$ units and so
$\gamma ^{-1} = 1$. The $\omega $ values used to mimic the time series
were obtained from stationary Feller
processes~\cite{fellerbib} with a $T_{i}$ relaxation for each $i$ equity
(see specific values of $\gamma \, T$
in Tab.~I). Looking to Fig.~\ref{pdf} we have observed that our
dynamical propose, using this simple approach, is
able to provide good probabilistic description of the data.

\section{Final remarks}
In this article we have analysed some statistical properties of the
traded volume equities that
constitute the DJ30 index, namely the dependence degree between time
series elements and stationary PDF.
For the dependence degree we have used a non-extensive generalised
Kullback-Leibler information
measure. With this procedure we have studied the dependence between
variables which decreases on a logarithmic way with
the lag. We have also verified that this decrease of the dependence is
much slower than the one observed in the correlation function.
This fact indicates that non-linearities are present in traded volume
dynamics and that they may be important factors in other
statistical features such as multi-fractality~\cite{jls-bariloche}.
Analysing the stationary distribution we have verified that it fits well
for a $q$-generalised inverted Gamma distribution presenting a
$q$ value around $\frac {6}{5}$ for all series. In addition, we
developed a dynamical mechanism which has as stationary PDF the
$q$-generalised inverted Gamma distribution. Further developments of
these model may be achieved using perturbative
calculus for a more accurate determination of $\gamma $~\cite{progress}
and determination of the ratio
between the scale of local relaxation and the mean traded volume
update~\cite{b-c-s}.

\bigskip

The authors thank C. Tsallis (particularly for the
remark~\cite{nota-ct}) and E.M.F. Curado
for their continuous encouragement and fruitful comments as well as F.D.
Nobre and C. Beck. Olsen
Data Services are acknowledged for have provided the data. This work was
done
under financial support from CNPq, PRONEX (Brazilian agencies) and
FCT/MCES (Portuguese agency).



\begin{references}
\bibitem{nota-ct} As a matter of fact it can be shown that:
$x^a e_q^{- \frac{x}{b}} = \Bigl[\frac{b}{q-1} \Bigr]^{\frac{1}{q-1}}
x^{a - \frac{1}{q-1}}\, e_q^{-\frac{b /
(q-1)^2}{x}}\;\;\;\;(q>1)$. In other words, by a simple redefinition of
parameters the inverted $q$-Gamma can be written
as the $q$-Gamma PDF.
\end{references}
\end{document}